\documentclass[aps,prmaterials,reprint,superscriptaddress,amsmath,amssymb,longbibliography]{revtex4-2}
\usepackage{graphicx}
\usepackage{xcolor}
\usepackage{textcomp}
\usepackage{upgreek}
\usepackage[colorlinks=true,linkcolor=blue,citecolor=blue,urlcolor=blue]{hyperref}

\newcommand{\degC}{\textdegree C}
\newcommand{\Tc}{T_{\mathrm{c}}}
\newcommand{\Bc}{B_{\mathrm{c}}}
\newcommand{\RN}{R_{\mathrm{N}}}

\begin{document}

\title{Solid-state reaction synthesis of superconducting refractory-metal germanides for Ge/SiGe quantum devices}

\author{Sebastiaan R. Roelofs}
\affiliation{{QuTech and Kavli Institute of Nanoscience, Delft University of Technology, Delft, The Netherlands}}
\author{Jan Cornelis Wolff}
\affiliation{{QuTech and Kavli Institute of Nanoscience, Delft University of Technology, Delft, The Netherlands}}
\author{Nick van Loo} 
\affiliation{{QuTech and Kavli Institute of Nanoscience, Delft University of Technology, Delft, The Netherlands}}
\author{Karina L. Hudson}
\affiliation{{QuTech and Kavli Institute of Nanoscience, Delft University of Technology, Delft, The Netherlands}}
\author{Giordano Scappucci}
\affiliation{{QuTech and Kavli Institute of Nanoscience, Delft University of Technology, Delft, The Netherlands}}
\author{Greg P. Mazur}
\email{gmazur@ucsb.edu}
\affiliation{Materials Department, University of California, Santa Barbara, California 93106, USA}
\affiliation{Department of Materials, University of Oxford, Parks Road, Oxford OX1 3PH, United Kingdom}
\date{\today}

\begin{abstract}
Solid-phase reactions offer a route to integrate superconductors into Ge/SiGe heterostructures, but forming high-quality superconducting layers near the buried Ge quantum well remains challenging. We investigate reactions of V, Ta, and Nb with elemental Ge and with the Si$_{0.2}$Ge$_{0.8}$ barrier of Ge/SiGe heterostructures at processing temperatures of 365--390\degC. Cross-sectional electron microscopy and low-temperature electrical transport show that  deposition on Ge at 390\degC \, in which approximately 2~nm metal increments are separated by growth stops for in-situ annealing, produces reactions throughout the deposited films and superconducting transitions at approximately 0.14~K for V--Ge, 2.1~K for Ta--Ge, and 2.8~K for Nb--Ge. The Nb--Ge layer has an out-of-plane critical field of approximately 0.8~T at 2.1~K. In the heterostructure samples, intermixing is confined to regions approximately 5--12~nm thick near the metal--SiGe interface, leaving a substantial residual metal layer. Microscopy and transport together suggest that this residual film dominates the measured superconducting response. These results establish low-temperature formation of superconducting refractory-metal--Ge layers while identifying the limited interfacial reaction as the principal obstacle to extending this approach toward the buried quantum well.
\end{abstract}

\maketitle

\section{Introduction}
Planar Ge/SiGe heterostructures have emerged as a promising platform for hole-spin qubits~\cite{Jirovec2021,Hendrickx2020,Hendrickx2021,Scappucci2021,Fang2023}. Holes in strained Ge combine large out-of-plane $g$ factors ($|g_z|\approx 10$--$20$)~\cite{Drichko2018,Hendrickx2024}, strong intrinsic spin--orbit coupling that enables electrical qubit control~\cite{Terrazos2021,Hendrickx2020}, mobilities exceeding $7\times10^{6}$~cm$^2$\,V$^{-1}$\,s$^{-1}$~\cite{Lodari2021,Myronov2023,Stehouwer2023,Costa2024,Myronov2026}, and a nuclear-spin-poor lattice that can be isotopically purified~\cite{Zwanenburg2013,Itoh2014,Assali2024}. These properties make Ge attractive not only for isolated spin qubits, but also for hybrid architectures in which semiconductor quantum devices are integrated with superconducting elements.

Superconductivity could extend the Ge platform in two complementary directions. Superconducting resonators can mediate long-range microwave coupling between spin qubits, as demonstrated in Si~\cite{Mi2018,Samkharadze2018,Borjans2020,HarveyCollard2022,Yu2023,Dijkema2025} and now pursued in Ge~\cite{Sagi2024,Nigro2024}. Proximity-induced pairing in the hole gas provides a route to Josephson devices, proximitized quantum dots, Andreev spin qubits, quantum-dot Kitaev chains, and topological superconductivity~\cite{Lutchyn2018,Prada2020,Hays2021,PitaVidal2023,Hoffman2025,Dvir2023,Laubscher2024a,Laubscher2024b,Lakic2025}. In such devices, the semiconductor--superconductor interface is a central material parameter, as it sets the strength and quality of the induced superconductivity~\cite{vanLoo2023}.

Most superconducting Ge devices so far have relied on metal deposition onto an etched SiGe top barrier~\cite{Vigneau2019,Ridderbos2020,Aggarwal2021}. Etching can damage the semiconductor surface and reduce interface transparency. A different strategy is to form the superconductor directly inside the heterostructure by a solid-phase reaction. Tosato \emph{et al.}\ demonstrated this approach with PtSiGe formed inside the SiGe barrier at 400\degC, obtaining a hard induced gap~\cite{Tosato2023}. PtSiGe has since been reproduced independently in planar Ge Josephson junctions with a comparable induced gap~\cite{Hinderling2024} and has enabled a proximitized quantum dot~\cite{Lakic2025}. PtSiGe itself, however, has a relatively low critical temperature, $\Tc\approx0.5$~K, and an in-plane critical field of approximately 0.4~T, limiting the accessible energy and magnetic-field scales. Partial substitution of Ir for Pt extends these scales, yielding a PtIrSiGe alloy with $\Tc\approx1.85$~K and an out-of-plane critical field of 1.9~T~\cite{Viswanathan2026} at 80\, mK. Solid-phase epitaxy of a broader set of group-VIII germanosilicides---including Pd-, Rh-, and Ir-based compounds---has been characterized by Li \emph{et al.}, who found IrGe superconducting at up to 3.4~K~\cite{Li2026}.

Refractory-metal germanides offer a further route to higher energy and magnetic-field scales. V, Ta, and Nb form several superconducting germanide and silicide phases~\cite{Matthias1963,Ghosh1977}, including the A15 compounds Nb$_3$Ge and V$_3$Si, which belong to the same structural family as Nb$_3$Sn and have critical temperatures up to approximately 23~K and high upper critical fields~\cite{Hardy1954,Gavaler1973,Testardi1975,Stewart2015,Cardwell2022}. These phases are normally synthesized at temperatures above 700\degC\ or from stoichiometric sputter targets~\cite{Terada1984,Stewart2015}. In strained Ge/SiGe heterostructures, by contrast, the quantum well and top barrier are grown near 500\degC~\cite{Stehouwer2023}, which places a practical upper bound on subsequent thermal processing if the quantum well is to be preserved. It is therefore not known whether useful superconducting refractory-metal germanides can be formed within the heterostructure thermal budget, or whether these metals can react sufficiently far into the SiGe barrier to approach the buried quantum well.

Here we address these two questions by comparing V, Ta, and Nb deposited at 365--390\degC\ onto bulk Ge and onto the Si$_{0.2}$Ge$_{0.8}$ top barrier of Ge/SiGe heterostructures. We combine cross-sectional scanning transmission electron microscopy with low-temperature resistance measurement. On Ge, all three metals react through the deposited layer and form superconducting germanides, with Ta--Ge and Nb--Ge reaching $\Tc=2.1$ and 2.8~K, respectively. On the heterostructures, the same thermal budget produces only a thin reacted region at the metal--SiGe interface. The contrast shows that low-temperature germanide formation is feasible, but that reaction kinetics in the SiGe barrier currently limit extension of the process toward the quantum well.

\section{Experimental methods}\label{sec:methods}
\subsection{Heterostructures and substrates}
The heterostructures, grown on a Si substrate, consist of a strained Ge quantum well (16~nm) buried under a 55 nm thick Si$_{0.2}$Ge$_{0.8}$ top barrier~\cite{Lodari2021}. Since the barriers and quantum well are grown at 500\degC~\cite{Stehouwer2023}, we use this temperature as the upper thermal budget for all subsequent processing. Samples of bulk undoped Ge (001) wafers were processed in parallel to separate the metal--Ge reaction from the additional chemical and structural complexity of the SiGe barrier.

\subsection{Surface preparation}
Native GeO$_x$ was removed from the Ge samples in deionized water at 90\degC\ for 10~min~\cite{Strohbeen2024}. The native SiO$_x$ on the heterostructures was etched in HF:H$_2$O (1:7) for 30~s, after which the samples were transferred under isopropanol. Both substrate types were then dried with N$_2$ and loaded into the evaporator load lock within 1--2~min.

\subsection{Metal deposition and annealing}
V, Ta, and Nb films with total thicknesses of 30--40~nm were deposited by electron-beam evaporation in an ultra-high-vacuum chamber with a pressure of $10^{-9}$--$10^{-7}$~mbar. The substrates were kept at a fixed temperature using a heater directly below the sample, regulated by a thermocouple, with a maximum temperature of 390\degC. Most films were deposited step-wise: approximately 2~nm of metal was deposited at 0.05--0.1~\AA\,s$^{-1}$ and followed by a 5--15~min anneal at the growth temperature before the next increment was added~\cite{Strohbeen2024}. In continuous-growth samples, the full metal thickness was deposited at 0.05~\AA\,s$^{-1}$ and annealed only after deposition. Unless stated otherwise, the samples described in the main text were prepared by the step-wise process. 

After the final anneal, the films were capped to suppress oxidation. Most samples received 10~nm of AlO$_x$ deposited at 300\,K using e-beam evaporation. One Nb/SiGe sample instead received a 3~nm Ta cap, which can act as an oxygen-diffusion barrier~\cite{Kopas2021}.

\subsection{Transport measurements}
Unpatterned films were wire-bonded in a four-terminal in-line geometry and measured in a dilution refrigerator between 20~mK and several kelvin. A vector magnet provided fields up to 1~T perpendicular to the film and 5.5~T in plane. The resistance was measured using standard lock-in techniques and low-noise differential amplification with excitation currents of 0.1--1~$\upmu$A. We define $\Tc$ as the temperature at which the resistance falls to $0.1\RN$, where $\RN$ is the normal-state resistance. The out-of-plane critical field $\Bc$ is defined using the same 10\% criterion at fixed temperature.

\subsection{Electron microscopy}
Cross-sectional lamellae were prepared by focused-ion-beam milling (FIB) in a Helios G4 CX, using a Pt or W capping layer to protect the sample and placed on a Cu support. High-angle annular dark-field scanning transmission electron microscopy (HAADF-STEM) and energy-dispersive X-ray spectroscopy (EDX) were performed in a $C_{\mathrm{s}}$-corrected FEI Titan Cubed TEM operated at 300~keV. The cross-sections were exposed to air before microscopy and therefore contain a surface oxide that is not part of the as-grown buried interface. Atomic compositions were quantified from EDX spectra (excluding C and Cu) and carry an uncertainty of roughly 10~\% for the transition metals and Ge, while for the remaining elements it is $<2$\%. The uncertainty has not been plotted for clarity of the data. The protective capping layer materials for FIB Pt/W and Al in the AlO$_x$ capping are not shown, but are accounted for in the atomic fraction results below.  %

\begin{table*}[t]
\caption{Samples and superconducting properties. Depositions were performed at 365--390\degC. Step-wise (step) samples were annealed for 5--15~min after every approximately 2~nm increment, whereas continuous (cont.) samples were annealed after deposition while cooling down to room temperature. $\Tc$ is defined by $R=0.1\RN$, and $\Bc$ is the out-of-plane critical field at the stated temperature using the same resistance criterion. Compositions marked $^{\ast}$ require verification. Reacted layers listed as NA were not subjected to TEM.} 
\label{tab:samples}
\begin{ruledtabular}
\begin{tabular}{lllllllllll}
Metal & Substrate & Thickness & Growth & Temperature & Anneal & Cap & Reacted layer & $\Tc$ & $\Bc$ \\
      &           & (nm)      &        & (\degC)     & (min)  &     &               & (K)   & (T) \\
\hline
V  & Ge   & $15\times2$ & step  & 390 & $15\times5$  & AlO$_x$ & complete, V:Ge $\approx 1{:}1^{\ast}$     & 0.14 & Not measurable \\ %
V  & Ge   & 40          & cont. & 365 & --           & AlO$_x$ & NA                                        & 4.0  & 0.82 (2.3~K)\\ %
V  & SiGe & 40          & cont. & 365 & --           & AlO$_x$ & $\approx 7$~nm                            & 4.15 & 0.16 (3.8~K) \\ %
\hline
Ta & Ge   & $15\times2$ & step  & 390 & $15\times15$ & AlO$_x$ & complete, Ta:Ge $\approx 2{:}3^{\ast}$    & 2.1  & 0.19 (1.8~K) \\ %
Ta & SiGe & $15\times2$ & step  & 390 & $15\times15$ & AlO$_x$ & $\approx5$--10~nm                                  & 1.04 & 0.31 (0.8~K) \\ %
\hline
Nb & Ge   & $20\times2$ & step  & 390 & $20\times5$  & AlO$_x$ & complete, Nb:Ge $\approx 5{:}4^{\ast}$    & 2.8  & 0.77 (2.1~K) \\ %
Nb & SiGe & $15\times2$ & step  & 390 & $15\times5$  & Ta      & $\approx 7+5$~nm                          & 4.7  & 0.65 (4.0~K) \\ %
\end{tabular}
\end{ruledtabular}
\end{table*}

\section{Results}
Table~\ref{tab:samples} summarizes the samples and their superconducting properties. Our central comparison is between reactions on elemental Ge and on the Si$_{0.2}$Ge$_{0.8}$ top barrier of the heterostructure. For each material, we first establish whether the chosen thermal budget is sufficient to form a reacted superconducting layer on Ge and then apply a comparable recipe to the heterostructure. This side-by-side approach separates the intrinsic ability of the metal to form a superconducting germanide from the additional kinetic constraints introduced by the SiGe alloy.

\subsection{Low-temperature formation of superconducting germanides on Ge}
We first test whether V, Ta, and Nb can react with Ge at temperatures below the 500\degC\ heterostructure growth temperature. All three metals show complete reaction through the deposited film under at least one of the processing conditions studied here, demonstrating that superconducting refractory-metal germanides can be formed at 365--390\degC.

\begin{figure*}
\includegraphics[width=\textwidth]{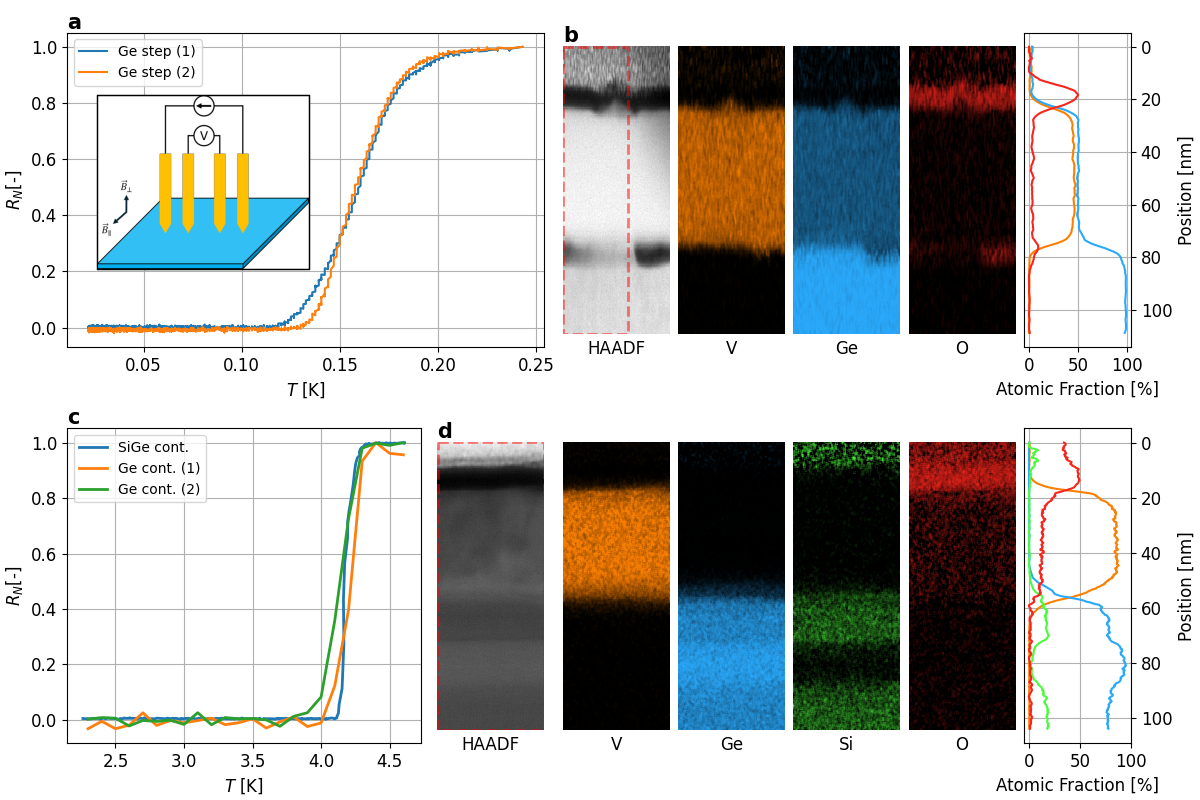}
\caption{Vanadium on Ge and on the Ge/SiGe heterostructure. (a) Resistance normalized to the normal-state value $\RN$ as a function of temperature for V deposited step-wise on Ge ($15\times2$~nm, 390\degC, 5~min anneals), measured at two positions on the film. The transition reaches $R=0.1\RN$ at 135~mK. The inset shows the four-terminal geometry and the definitions of $B_\perp$ and $B_\parallel$. (b) Cross-sectional HAADF-STEM image, EDX maps of V, Ge, and O, and corresponding atomic-fraction depth profile for the same sample. The red outline in the HAADF image marks the data used for the atomic fraction profile. V and Ge occupy the reacted layer between the AlO$_x$ cap and the Ge substrate. Residual oxygen near the original Ge surface marks the approximate initial interface, while voids appear immediately below it on the Ge side. (c) $R/\RN$ as a function of temperature for 40~nm V deposited continuously at 365\degC\ on the Ge/SiGe heterostructure and, using the same deposition conditions, on Ge at two positions. All three transitions occur between 4.0 and 4.15~K. (d) HAADF-STEM image, EDX maps of V, Ge, Si, and O, and depth profile of the heterostructure sample. The V film intermixes with the SiGe barrier only over approximately 7~nm near the interface.}
\label{fig:V}
\end{figure*}

For V, the superconducting target phases include the A15 compounds V$_3$Si ($\Tc\approx17$~K)~\cite{Hardy1954}, V$_3$Ge (6.2~K)~\cite{Ahmed2022}, and the pseudobinary V$_3$Si$_x$Ge$_{1-x}$ system~\cite{Pendrys1976}. These phases are normally prepared at substantially higher temperatures, and their superconducting properties are sensitive to stoichiometry and crystal structure. Figure~\ref{fig:V}(a,b) shows a V/Ge sample prepared by depositing $15\times2$~nm of V step-wise at 390\degC\ with 5~min waiting times between the steps. EDX reveals that the deposited V is incorporated throughout a reacted V--Ge layer with an approximate V:Ge ratio of 1:1. This composition does not correspond to the equilibrium compounds V$_3$Ge, V$_5$Ge$_3$, V$_{11}$Ge$_8$, or V$_{17}$Ge$_{31}$~\cite{Massalski1990}, suggesting a disordered or multiphase reaction product rather than a well-defined A15 phase. The layer becomes superconducting at $\Tc=140$~mK, with nearly identical transitions measured at two positions on the sample [Fig.~\ref{fig:V}(a)]. Both the composition and the transition temperature therefore argue against formation of V$_3$Ge.

The V/Ge cross-section also provides information about the reaction mechanism. A residual oxygen signal remains near the depth of the original Ge surface, while voids are visible directly below this marker plane on the Ge side [Fig.~\ref{fig:V}(b)]. We interpret this combination as evidence for a Kirkendall-type reaction in which Ge is the dominant diffusing species. In this picture, Ge diffuses into the V layer faster than V diffuses into the substrate, producing a net vacancy flux toward the Ge side of the original interface where vacancies can condense into voids. A similar diffusion asymmetry is known for refractory-metal disilicides~\cite{dHeurle1986,Nicolet1983}. Etch-induced pits cannot be excluded entirely, but their coincidence with the marker plane would be less natural.

\begin{figure*}
\includegraphics[width=\textwidth]{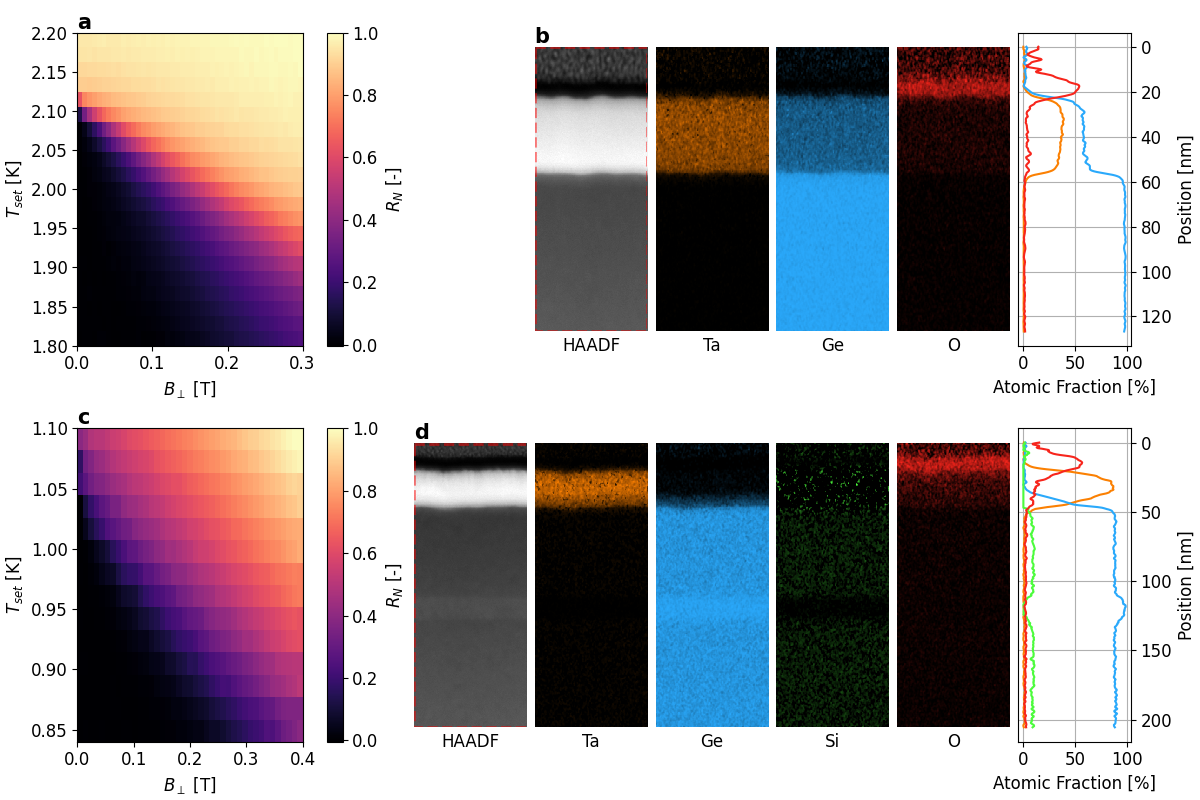}
\caption{Tantalum on Ge and on the heterostructure. (a) Resistance as a function of out-of-plane magnetic field $B_\perp$ and set temperature $T_{\mathrm{set}}$ for Ta deposited step-wise on Ge ($15\times2$~nm). At zero field, the resistance transition occurs at 2.1~K. (b) Cross-sectional HAADF-STEM image, EDX maps of Ta, Ge, and O, and the corresponding depth profile. The deposited Ta is incorporated into an approximately 40~nm reacted region below the AlO$_x$ cap.  (c) Same measurement as in (a) for the corresponding Ta process on the Ge/SiGe heterostructure. The zero-field transition is at 1.04~K and superconductivity persists to approximately 0.31~T at 0.8~K. (d) HAADF-STEM image, EDX maps of Ta, Ge, Si, and O, and the corresponding depth profile for the heterostructure sample. Most of the Ta remains in the metallic film, while only a 5--10~nm region near the interface is intermixed with the barrier. The Ge quantum well appears as a Si-free, Ge-rich band approximately 60~nm below the interface. Both samples carry an AlO$_x$ cap.}
\label{fig:Ta}
\end{figure*}

Ta provides a second test of the low-temperature reaction route. Relevant superconducting phases include TaSi$_2$ ($\Tc=0.35$~K)~\cite{Gottlieb1992}, TaGe$_2$ (approximately 1.9--2.7~K depending on growth method)~\cite{Li2024,Strohbeen2024,Ghosh1977}, and metastable A15 Ta$_3$Ge ($\Tc\approx8$~K)~\cite{Hanak1970}. Following the step-wise approach of Strohbeen \emph{et al.}~\cite{Strohbeen2024}, we deposited $15\times2$~nm of Ta on Ge. EDX indicates that the Ta is fully incorporated into a reacted layer [Fig.~\ref{fig:Ta}(b)]. The current quantification gives a Ta:Ge ratio of approximately 2:3, which lies between the equilibrium compounds Ta$_5$Ge$_3$ and TaGe$_2$~\cite{Massalski1990}. This ratio depends on the chosen X-ray lines, since the Ta-L lines overlap with the Ge-K lines; we therefore quantified the atomic fractions using the Ta-M and Ge-K lines. More generally, the reported composition of Ta--Ge films depends strongly on the measurement technique. For Ta deposited on Ge(001) at 400\degC, Strohbeen \emph{et al.} obtained approximately 38.5\% Ta from EDS, close to our value, whereas Rutherford backscattering spectrometry (RBS) on the same film gave a Ta-rich Ta:Ge ratio of 58.5:41.5, and XPS depth profiling likewise indicated a Ta-rich film~\cite{Strohbeen2024}. They attributed the Ta deficit in EDS to oxidation during sample transfer, overlapping line shapes, and the absence of a suitable standard, adopted the RBS value, and interpreted it using the phase diagram as a mixture of predominantly Ta$_5$Ge$_3$ with some TaGe$_2$~\cite{Strohbeen2024}. Their films showed no Ta--Ge diffraction peaks, and our TEM likewise did not identify a crystalline phase in the Ta or Nb samples, so in neither case can the phase be assigned structurally. The discrepancy of about 20 percentage points between EDS and RBS in Ref.~\cite{Strohbeen2024} exceeds our quoted EDX uncertainty, and our 2:3 ratio may therefore underestimate the Ta content. A standard-free technique such as RBS, or atom probe tomography, would be needed to establish the absolute composition. 

The Ta--Ge layer becomes superconducting at $\Tc=2.1$~K [Fig.~\ref{fig:Ta}(a)]. At 1.8~K, superconductivity persists to an out-of-plane field of order 0.19~T using the present resistance criterion. The transition temperature is consistent with reported TaGe$_2$ films and crystals~\cite{Li2024,Strohbeen2024}, although the present microscopy does not yet establish that this is the only phase present.

\begin{figure*}
\includegraphics[width=\textwidth]{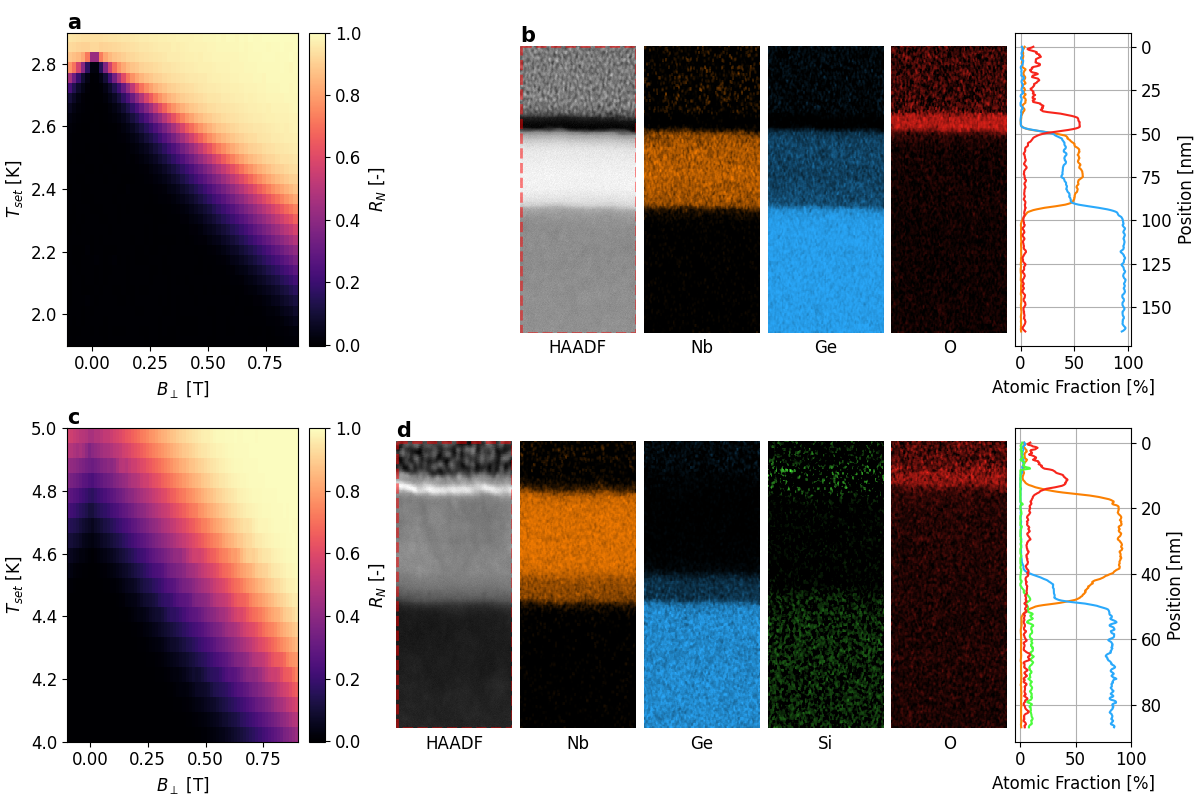}
\caption{Niobium on Ge and on the heterostructure. (a) Resistance as a function of $B_\perp$ and $T_{\mathrm{set}}$ for Nb deposited step-wise on Ge ($20\times2$~nm, AlO$_x$ cap). The zero-field transition occurs at 2.8~K and the superconducting region extends to approximately 0.77~T near 2.1~K. (b) Cross-sectional HAADF-STEM image, EDX maps of Nb, Ge, and O, and the corresponding depth profile. The deposited Nb is incorporated into an approximately 50~nm reacted Nb--Ge layer beneath the AlO$_x$ cap. (c) Same transport measurement for Nb deposited step-wise on the Ge/SiGe heterostructure ($15\times2$~nm, Ta cap). (d) Cross-sectional HAADF-STEM image, EDX maps of Nb, Ge, Si, and O, and the corresponding depth profile for the heterostructure sample. Beneath the Ta cap, the Nb film reacts with the SiGe barrier over approximately 12~nm, forming an approximately 7~nm Nb--Ge-rich region followed by an approximately 5~nm Nb--Si--Ge interfacial region.} %
\label{fig:Nb}
\end{figure*}

Among the three materials, Nb gives the most favourable superconducting properties on Ge. Nb forms high-$\Tc$ A15 compounds Nb$_3$(Ge,Si)~\cite{Gavaler1973,Terada1984}, as well as lower-temperature phases such as chiral C40 NbGe$_2$ ($\Tc=2.06$~K)~\cite{Emmanouilidou2020} and NbSi$_2$ ($\Tc=0.13$~K)~\cite{Lasjaunias1993}. Because the A15 phases generally require much higher formation temperatures, the low-temperature process is more naturally expected to produce a non-A15 germanide. Step-wise deposition of $20\times2$~nm Nb on Ge produces a fully reacted layer [Fig.~\ref{fig:Nb}(b)] with an approximate Nb:Ge ratio of 5:4. This composition lies between the equilibrium compounds Nb$_5$Ge$_3$ and NbGe$_2$~\cite{Massalski1990}, and may therefore represent a disordered or multiphase layer rather than a single equilibrium phase.

The reacted Nb--Ge layer has $\Tc=2.8$~K [Fig.~\ref{fig:Nb}(a)], exceeding the reported bulk transition temperature of NbGe$_2$~\cite{Emmanouilidou2020}. At 2.1~K, the layer remains superconducting to an out-of-plane field of approximately 0.77~T, the largest field scale observed among the reacted Ge samples. Taken together, the V, Ta, and Nb results establish that the 365--390\degC\ processing window is sufficient to form superconducting refractory-metal germanides on elemental Ge, even though the resulting compositions need not correspond to the equilibrium high-$\Tc$ phases.

\subsection{Reaction is confined near the SiGe interface}
Next, we apply comparable processing conditions to the Si$_{0.2}$Ge$_{0.8}$ top barrier. In all three material systems, the result differs qualitatively from the corresponding Ge sample: reaction is confined to a narrow interfacial region and a substantial fraction of the deposited metal remains unreacted. The transport response must therefore be interpreted together with the microscopy rather than assigned directly to the reacted layer.

For V, 40~nm deposited continuously at 365\degC\ intermixes with the SiGe barrier only within approximately 7~nm of the interface [Fig.~\ref{fig:V}(d)]. The film shows a transition at $\Tc=4.15$~K [Fig.~\ref{fig:V}(c)], which is more likely to be a remaining metallic V film, with a modest suppression that could arise from reduced thickness or disorder~\cite{Kuzmenko1975}. A V film deposited continuously on Ge using the same 365\degC\ parameters shows nearly the same transition at $\Tc=4.0$~K [Fig.~\ref{fig:V}(c)], suggesting that continuous growth followed by a single anneal is insufficient to drive the strong reaction observed for the step-wise V/Ge sample. The lower growth temperature used for the continuous films may also contribute. However, this sample was not examined by TEM, so we cannot distinguish an essentially unreacted V film from a different reaction product with a similar $\Tc$, such as a more V-rich germanide. Because the fully reacted V--Ge layer itself has a much lower $\Tc$, we did not pursue V as the leading candidate for proximity devices. 

For Ta, the step-wise process that reacts throughout the deposited film on Ge produces only a 5--10~nm intermixed layer on the heterostructure [Fig.~\ref{fig:Ta}(d)]. This region remains far above the quantum well. The sample has $\Tc=1.04$~K and $\Bc=310$~mT at 800~mK [Fig.~\ref{fig:Ta}(c)]. These values are closer to the superconducting properties of disordered $\beta$-Ta, for which $\Tc$ is approximately 0.5~K~\cite{Read1965}, than to those of $\alpha$-Ta, which has $\Tc\approx4.4$~K~\cite{Roberts1976}. We therefore attribute the measured transport properties primarily to the remaining Ta film.

Nb shows the deepest intermixing of the three metals. In the Nb/SiGe sample with a Ta cap, EDX resolves approximately 7~nm of Nb--Ge-rich material above an approximately 5~nm Nb--Si--Ge region [Fig.~\ref{fig:Nb}(d)]. This compositional plateau followed by a transition region is distinct from the simpler interfacial profiles observed for V and Ta. The sample has $\Tc=4.7$~K and $\Bc=0.65$~T at 4~K [Fig.~\ref{fig:Nb}(c)]. The TEM shows that roughly 20~nm of Nb remains above approximately 12~nm of reacted Nb--(Si)Ge. The most conservative interpretation is that the observed superconductivity is dominated by the remaining Nb film, possibly modified by proximity coupling to the reacted layer~\cite{Langa2024}. A spatial distribution of local transition temperatures could further broaden the measured transition. We don't expect thin Ta cap to contribute to superconducting properties of the measured film. Critical temperature of  $\alpha$-Ta is around 4.4K for crystalline thick films grows at high temperatures. Despite the material-dependent details, the common result is that the reaction stalls near the metal--SiGe interface. 

\section{Discussion}
The side-by-side comparison isolates the main limitation of the materials. At 365--390\degC, V, Ta, and Nb can all react extensively with elemental Ge, yet the same processes penetrate only a few nanometers into Si$_{0.2}$Ge$_{0.8}$. The difference is therefore not simply whether a given metal is capable of forming a germanide at this temperature; it is the kinetics of reaction with the alloy barrier.

Several mechanisms could contribute. The Ge wafers and epitaxial SiGe barriers are both crystalline, but their defect densities are unlikely to be identical. A higher density of defects in the bulk Ge wafers could accelerate diffusion relative to the high-quality epitaxial barrier. The SiGe alloy also changes the bonding environment relative to elemental Ge, and forming a reaction product requires redistribution of both Si and Ge. Thin-film formation temperatures provide an empirical indication of the corresponding kinetic barriers: VSi$_2$ on Si forms only near 570--650\degC\ with an activation energy of approximately 1.7~eV, while V$_3$Si forms above approximately 730\degC~\cite{Krautle1974}; V, Nb, and Ta germanides likewise require comparatively high temperatures in conventional thin-film reactions~\cite{Gaudet2006}. 

Strain and point-defect physics may further modify the reaction. Kringh{\o}j \emph{et al.}\ showed that vacancy-mediated Sb diffusion in Si$_{0.91}$Ge$_{0.09}$ is enhanced by compressive strain and retarded by tensile strain~\cite{Kringhoj1996}. Self-diffusion in Ge is vacancy-mediated, whereas both vacancies and interstitials can contribute in Si~\cite{Bracht2011,Chroneos2014}. The diffusivities of V, Ta, and Nb in Si$_{0.2}$Ge$_{0.8}$ are therefore unlikely to follow a simple interpolation between pure Si and pure Ge and require dedicated study.

The V/Ge microstructure adds another constraint on how far the reaction should ultimately be driven. The voids located on the Ge side of the apparent marker plane are consistent with Ge being the dominant moving species. If the same mechanism operates in the SiGe barrier, extending the reacted layer toward the quantum well would inject vacancies into the semiconductor side of the interface. The reaction should therefore be advanced only as far as required to establish strong superconducting coupling, because excessive reaction could introduce vacancy-related disorder near the quantum well.

The comparison with PtSiGe is particularly informative~\cite{Tosato2023}. Firstly, Pt was deposited at room temperature and subsequently reacted by rapid thermal annealing at 400\degC, whereas in the present work V, Ta, and Nb were deposited directly onto hot substrates to promote intermixing. The annealing method may itself play a role: rapid thermal annealing is carried out in an inert-gas ambient, such as Ar~\cite{Viswanathan2026}, with fast heating and cooling ramps, whereas our films were annealed in situ in ultra-high vacuum. Differences in ambient, thermal history, and temperature calibration between the two approaches could therefore affect the reaction. Secondly, the thin-film formation temperature of PtGe is around 380--440\degC~\cite{Gaudet2006}, and PtSiGe can be formed near 400\degC~\cite{Tosato2023}, while refractory-metal silicides and germanides generally require substantially higher temperatures. Therefore, the simplest interpretation is that V, Ta, and Nb diffuse too slowly through the SiGe barrier on the timescales used in this work. The barrier thickness may compound this difference: the 55~nm top barrier used here is more than twice as thick as the 22~nm barrier in Ref.~\cite{Tosato2023}. %

These observations suggest three practical routes forward. First, the processing temperature can be raised to the 500\degC\ heterostructure growth temperature: PtIrSiGe formed by rapid thermal annealing at 500\degC\ proximitizes the buried Ge quantum well, and TEM shows that the quantum well remains structurally intact~\cite{Viswanathan2026}. Second, substantially longer post-deposition anneals may increase the reaction depth without requiring a larger metal thickness; at present, adding more metal mainly increases the amount of unreacted film. Third, a thinner top barrier would reduce the distance between the reacted region and the quantum well, at the possible cost of reduced mobility. %

\section{Conclusion}
Step-wise deposition at 390\degC\ produces superconducting reaction layers throughout the deposited V, Ta, and Nb films on elemental Ge. The resulting transition temperatures are approximately 0.14~K for V--Ge, 2.1~K for Ta--Ge, and 2.8~K for Nb--Ge. The Nb--Ge layer has an out-of-plane  critical field of approximately 0.8~T at 2.1~K.

In the Ge/SiGe samples, intermixing is confined to regions approximately 5--12~nm thick near the metal--SiGe interface. Microscopy and transport suggest that the residual metal films dominate the measured superconductivity. The results identify limited interfacial reaction as the principal processing obstacle, without establishing a unique microscopic cause. They separate two requirements for Ge/SiGe integration: low-temperature formation of superconducting reaction layers, demonstrated here on elemental Ge, and controlled extension of those layers toward the buried quantum well while preserving its quality, which remains unresolved under the conditions investigated.

\begin{acknowledgments}
For the Titan TEM results, we acknowledge support from the Kavli Institute of Nanoscience, Delft University of Technology, and the Netherlands Electron Microscopy Infrastructure (NEMI), project number 184.034.014, part of the National Roadmap and financed by the Dutch Research Council (NWO). 
\end{acknowledgments}

\section*{Data availability}
The scripts to produce the figures, raw transport measurement data and TEM data used in this work are available on Zenodo \cite{Roelofs2026}.

\bibliography{refs}

\end{document}